\documentstyle[12pt]{article}
\newcommand{\be}{\begin{equation}}
\newcommand{\ee}{\end{equation}}
\newcommand{\ba}{\begin{eqnarray}}
\newcommand{\ea}{\end{eqnarray}}

\begin{document}
\setlength{\baselineskip}{.7cm}
\renewcommand{\thefootnote}{\fnsymbol{footnote}}
\newcommand{\lp}{\left(}
\newcommand{\rp}{\right)}
\renewcommand{\thefootnote}{\fnsymbol{footnote}}

\sloppy

\begin{center}
\centering{\bf \Large Stock market crashes are outliers}
\end{center}

\begin{center}
\centering{Anders Johansen $^1$ and Didier Sornette $^{2,3}$ \\

{\it $^1$ CATS, Niels Bohr Institute, Blegdamsvej 17, DK-2100, Denmark

$^2$ Department of Earth and Space Science and Institute of Geophysics and \\
Planetary Physics University of California, Los Angeles, California 90095\\

$^3$ Laboratoire de Physique de la Mati\`ere Condens\'ee,
CNRS UMR6622\\ Universit\'e de Nice-Sophia Antipolis, B.P. 71, Parc
Valrose, 06108 Nice Cedex 2}
}
\end{center}
\vskip 4cm
{\bf Abstract} \\
We call attention against what seems to be a widely held misconception
according to which large crashes are the largest events of distributions
of price variations with fat tails. We demonstrate on the Dow Jones Industrial
Average that with high probability the three largest crashes in this
century are outliers.
This result supports the suggestion that large crashes result from specific
amplification processes that might lead to observable pre-cursory signatures.

\vskip 3cm
PACS numbers\,: 01.75+m ; 02.50+s ; 89.90+n

\pagebreak

Stock markets can exhibit very large motions, such as rallies and crashes.
These are the most extreme deviations from the ingrained Gaussian description
that was first shaken by Mandelbrot
(see \cite{Mandelbrot} and references therein) when he proposed to use
L\'evy distributions.
L\'evy distributions are characterised by a fat tail decaying as a power
law with index
between $0$ and $2$. Recently, physicists have characterised more precisely
the distribution
of market price variations [2-4] and found that
a power law truncated by an exponential provides a reasonable fit at short
time scales (less
than one day), while at larger time scales the distributions cross over
progressively to
the Gaussian distribution which becomes approximately correct for monthly
and larger scale price variations. Alternative representations exist
\cite{Ghashghaie}
corresponding to different models inspired from an analogy with turbulence.
These two classes of descriptions can only be distinguished using
higher order statistics \cite{muzy}.

This has led naturally to the idea that the stock market could exhibit
self-organising behaviour [7-10] where
large motions can occur relatively often in contrast to
what is expected within the Gaussian description.

The purpose of this short note is to point out the shortcomings of this
concept when applied to the very largest crashes. We would like to stress
the danger of amalgamating the existence of large crashes with the existence
of a fat tail in the distribution of stock market prices. Our message is
that the largest crashes of this century are outliers.

This is suggested by figure \ref{newdd1}, which shows the number of times a
given level of draw down has occurred in the Dow Jones Industrial Average in
the period 1900-1994. A draw down is defined as the cumulative loss from the
last local maximum to the next local minimum.
The number of draw downs DD smaller than $\approx 15
\%$ is well fitted by an exponential law
\be
N(\mbox{DD}) = N_0 ~e^{-\mbox{DD}/\mbox{DD}_c}~,  ~~~~{\rm with} ~DD_c
\approx 1.8 \%~. \label{zae}
\ee
This exponential fall-off is compatible with the previous results [2-6]
because the  time scales is already in the cross-over regime converging
to the Gaussian distribution. The three largest events seem to be outside
the range of the exponential fit. These events are in chronological order:
World War 1, Wall Street 1929.8 and Wall Street 1987.8. The largest is the
crash of October 1987, then World War 1 and Wall Street 1929.8. To forestall
any criticism related to the influence of the binning can have on the
appearance
of such a fit, we have re-binned distribution of draw downs using a 4 times
larger bin, see figure \ref{newdd4}. We still see an exponential distribution
with approximately the same decay constant (DD$_c \approx 2.2$ \%) for all but
the $3$ largest crashes. Note that the decrease in the distance between the
crash of 1929 and the exponential fit is partly due to the fact that the
larger binning attributes a decrease of 22.0\% to the crash and not the correct
value of 23.6\%. Let us also mention that the draw
down have all approximately the same duration\,: the average duration is
$3.3$ days to be
compared with $3.2$ for the 6 crashes larger than $15$\%.

To quantify how much these three events deviates from (\ref{zae}), we can
calculate what would be, according to (\ref{zae}),
 the typical return time of a draw down of amplitude
equal to or larger than the second largest of $28.8 \%$. Expression
(\ref{zae}) gives the number of draw down
equal to or larger than DD. Then $N_0$
is simply the total number of drawdown larger than $1\%$ in
a century. The fit yields $N_0 = 2360$, which is not too far from
the exact number $2789$ of drawn downs larger than $1\%$.
Taking the largest decay constant $DD_c$ of the two fits to be conservative,
expression (\ref{zae}) predicts
the number of drawn downs equal to or larger than $28.8\%$ per century to be
$\approx 0.006$. The
return time of draw down equal to or larger than $28.8\%$ would then be the
number of centuries $n$
such that $0.006~n \sim 1$. This yields $n \sim 160$ centuries. Taking the
other value
$DD_c =1.8\%$ yields a return time of $3000$ centuries. In contrast, the market
has sustained two such events in less than a century.

As an additional test, we have used a more sophisticated null-hypothesis than
that of an exponential and generated $10.000$ surrogate data sets
corresponding to approximately {\em one million} years using a GARCH(1,1)
model estimated from the true index with a t-student distribution with four
degrees of freedom \cite{GARCH}.
Among these $10.000$ surrogate data sets
only two had $3$ draw downs above $22$\% and none had $4$. However, $3$ of
these $6$ ``crashes'' showed a rather abnormal behaviour in the sense that
they were preceded by a draw up of comparable size as the ``crash'' (for real
crashes, the reverse is often seen, {\it i.e.}, large crashes are typically
followed by large draw ups). This means that in a million years of
``Garch-trading'', with a reset every century, {\it never} did $3$ crashes
occur. Furthermore, none were preceded by log-periodic signatures as found
for the 1929 and 1987 crashes \cite{Sornette1,Sornette2,Feigen}.

Figure \ref{ud2} shows the number of times a given level of draw up
has occurred in the Dow Jones Industrial Average of the same period. Here,
a power law
\be
N(\mbox{UD}) = \biggl({\mbox{UD}^{(+)} \over \mbox{UD} }\biggl)^{1+\mu}~,
~~{\rm with}~~
\mu \approx 2.0 ~,
\label{zadfe}
\ee
accounts reasonably well for the data except for the very smallest events.
There are no outliers for the draw ups and the exponent is not unreasonably
far from the value around $1.5$ that is found at smaller time scales. However,
the difference is significant and could be interpreted as an effective
power law
characterising the cross-over from the truncated L\'evy law at
short time scales to the Gaussian distribution at long time scales.

Crashes have recently been modelled as special critical crises [11-17].
The underlying hypothesis is that stock market crashes are caused by the
slow buildup
of powerful subterranean forces that come together in one critical instant.
The use of the word ``critical'' is not purely literary here\,: in
mathematical
terms, complex dynamical systems such as the stock market can go through
so-called ``critical'' points, defined as the explosion to infinity of a
normally
well-behaved quantity.

Both crashes and the price variations with their fat tail distributions and
anomalous
correlations are thus proposed to be the result of endogenous
self-organising process, however
they are not the same phenomenon. There is rather a coexistence of
self-organisation and criticality. Here, self-organisation refers to
the globally stationary state of the market in normal times with fat tail
distributions. The criticality here
describes the special times when a great crash occurs which has been
documented to be preceded by
a rising susceptibility and pre-cursory signals in a way similar to a critical
instability. Such a coexistence between self-organisation and criticality
has been recently demonstrated in a hierarchical model of earthquakes
\cite{Huang},
in which a coexistence of self-organisation of the crust at large time
scales and a critical nature of large earthquakes were found. The critical 
nature of
the large
cascades emerges from the interplay between the long-range
correlations of the self-organised state and the hierarchical
structure of interactions\,: a given level of the hierarchical rupture is
like a critical point to all the lower levels, albeit with a finite size.
The finite size effects are thus intrinsic to the process.

The fact that large crashes are outliers implies that they are probably
triggered by additional
amplifying factors. An important consequence is that specific signatures
of their
presence could exist, as proposed in [11-17] , similar to
precursors before instabilities. This is in contrast to the almost total
absence of
precursors before a large avalanche which is NOT an outlier in a
self-organised state
\cite{Sammis}.

\newpage

\begin{figure}
\caption{\label{newdd1} Number of times a given level of draw down has been
observed in
this century in the Dow Jones Average. The bin-size is 1\%. A threshold of
1\% has been applied. The fit is equation (\protect\ref{zae}) with
$N_0 \approx 2360$ and DD$_c\approx 0.018$.}
\end{figure}

\newpage

\begin{figure}
\caption{\label{newdd4} Number of times a given level of draw down has been
observed in
this century in the Dow Jones Average. The bin-size is 4\%. A threshold of
1\% has been applied. The fit is equation (\protect\ref{zae}) with
$N_0 \approx 3358$ and DD$_c \approx 0.022$.}
\end{figure}

\newpage

\begin{figure}
\caption{\label{ud2} Number of times a given level of draw up has been
observed in
this century in the Dow Jones Average. A threshold of 1\% has been applied.
The fit is equation (\protect\ref{zadfe}) with $\mu \approx 2.0$.}
\end{figure}

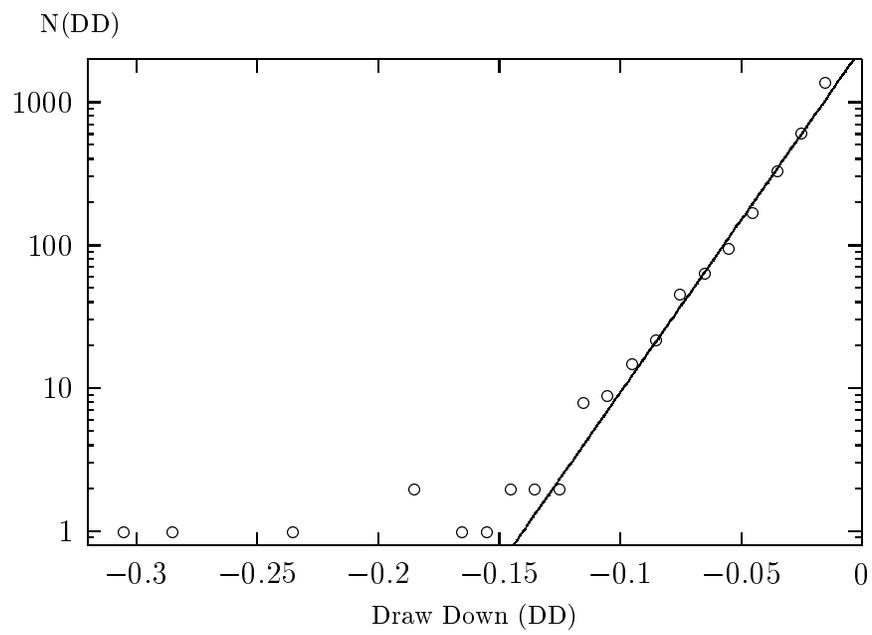
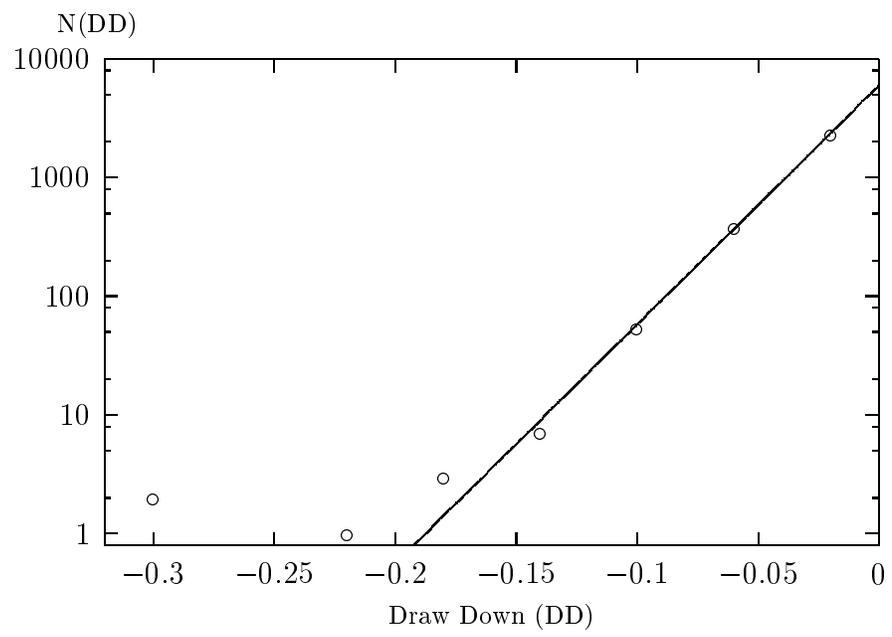
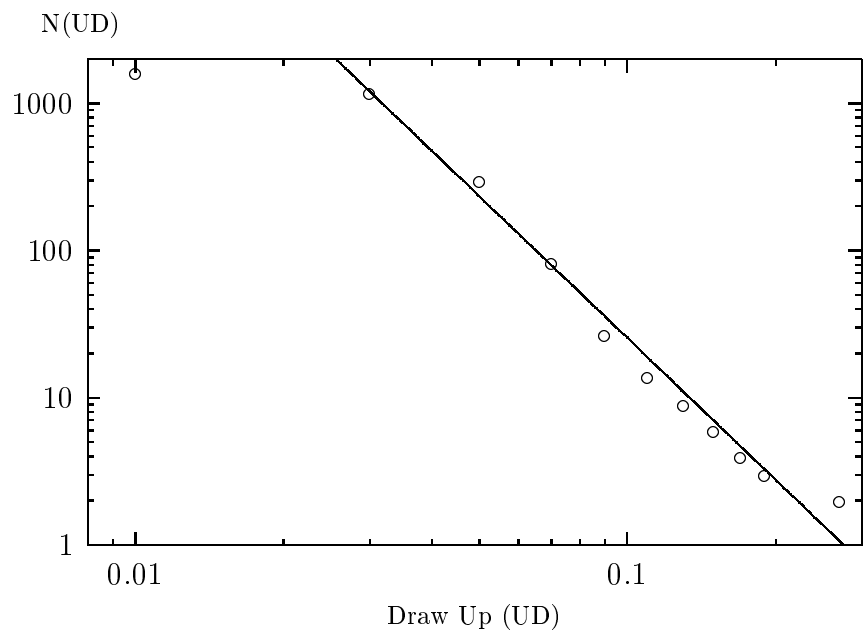
\end{document}